\documentclass[%
pr,superscriptaddress,
 amsmath,amssymb,
 reprint,%
]{revtex4-1}

\usepackage{graphicx}
\usepackage{dcolumn}
\usepackage{bm}
\usepackage[normalem]{ulem}

\usepackage{color}

\usepackage[unicode=true,colorlinks=true,citecolor=blue,urlcolor=blue]{hyperref}

\begin{document}

\preprint{AIP/123-QED}

\title{Spin-noise-based magnetometry of an $n$-doped GaAs microcavity in the field of elliptically polarized light  }



\author{I.~I.~Ryzhov}
\author{G.~G.~Kozlov}
\affiliation{Spin Optics Laboratory, St.-Petersburg State University, 1 Ul'anovskaya, Peterhof, St.-Petersburg 198504, Russia}



\author{D.~S. Smirnov}
\affiliation{Ioffe Institute of the RAS, 26 Polytekhnicheskaya, St.-Petersburg 194021, Russia}

\author{M.~M. Glazov}
\affiliation{Ioffe Institute of the RAS, 26 Polytekhnicheskaya, St.-Petersburg 194021, Russia}
\affiliation{Spin Optics Laboratory, St.-Petersburg State University, 1 Ul'anovskaya, Peterhof, St.-Petersburg 198504, Russia}

\author{Yu.~P.~Efimov}
\affiliation{Resource Center ``Nanophotonics'', St.-Petersburg State University, 1 Ul'anovskaya, Peterhof, St.-Petersburg 198504, Russia}

\author{S.~A.~Eliseev}
\affiliation{Resource Center ``Nanophotonics'', St.-Petersburg State University, 1 Ul'anovskaya, Peterhof, St.-Petersburg 198504, Russia}

\author{V.~A.~Lovtcius}
\affiliation{Resource Center ``Nanophotonics'', St.-Petersburg State University, 1 Ul'anovskaya, Peterhof, St.-Petersburg 198504, Russia}

\author{V.~V.~Petrov}
\affiliation{Resource Center ``Nanophotonics'', St.-Petersburg State University, 1 Ul'anovskaya, Peterhof, St.-Petersburg 198504, Russia}

\author{K.~V. Kavokin}
\affiliation{Ioffe Institute of the RAS, 26 Polytekhnicheskaya, St.-Petersburg 194021, Russia}
\affiliation{Spin Optics Laboratory, St.-Petersburg State University, 1 Ul'anovskaya, Peterhof, St.-Petersburg 198504, Russia}

\author{A.~V.~Kavokin}
\affiliation{Department of Physics \& Astronomy,
	University of Southampton, Southampton SO17 1BJ, United Kingdom}
\affiliation{Spin Optics Laboratory, St.-Petersburg State University, 1 Ul'anovskaya, Peterhof, St.-Petersburg 198504, Russia}

\author{V.~S.~Zapasskii}
\affiliation{Spin Optics Laboratory, St.-Petersburg State University, 1 Ul'anovskaya, Peterhof, St.-Petersburg 198504, Russia}

\begin{abstract}
	Recently reported optical nuclear orientation in the $n$-doped GaAs microcavity under pumping in nominal transparency region of the crystal [Appl. Phys. Lett. {\bf 106}, 242405 (2015)]  has arisen a number of questions, the main of them concerning  mechanisms of angular momentum transfer from the light to the nuclear spin system and the nature of the light-related magnetic fields accompanying the optical nuclear polarization. In this paper, we use the spin noise spectroscopy for magnetometric purposes, particularly, to study effective fields acting upon electron spin system of an $n$-GaAs layer inside a high-Q microcavity in the presence of elliptically polarized probe beam. In addition to the external magnetic field applied to the sample in the Voigt geometry and the Overhauser  field created by optically oriented nuclei, the spin noise  spectrum reveals an additional effective, ``optical'', magnetic field produced by elliptically polarized probe itself. This field is directed along the light propagation axis, with its sign being  determined by the sign of the probe helicity and its magnitude depending on degree of circular polarization and intensity of the probe beam. We analyze properties of this optical magnetic field and suggest that it results from the optical Stark effect in the field of the elliptically polarized electromagnetic wave.
\end{abstract}

\maketitle

%

\section{Introduction}\label{sec:intro}

Spin noise spectroscopy (SNS), primarily demonstrated on atomic systems \cite{Zap,Mitsui,Crooker}, has recently gained popularity mainly due to its application to semiconductor structures, with respect to which this technique proved to be most efficient~\cite{Oest,Os}. The basis of the SNS is provided by the fluctuation-dissipation theorem, which implies possibility of detecting resonances of linear susceptibility of the medium without its excitation, by ``listening'' to a noise of the medium in its equilibrium state. As applied to magnetic resonance spectroscopy, this possibility can be realized by detecting fluctuations of the Faraday rotation for a probe beam passing through a transparent (at the  probe wave frequency) paramagnet in a magnetic field directed across the light beam propagation direction. This  fluctuations are proportional to the fluctuations of the media magnetization. In this configuration, the Faraday rotation noise spectrum reveals a peak at the magnetic resonance  frequency corresponding to the precession of spontaneous fluctuations of the spin ensemble at the Larmor frequency. Since this technique, unlike conventional electron spin resonance (ESR) spectroscopy, does not imply an excitation of the resonance, it was considered to be essentially nonperturbative. Along with this property, which was initially regarded as the most important merit of SNS, the new technique has revealed a number of capabilities inaccessible either to conventional ESR spectroscopy, or even to linear optical spectroscopy in general (see, e.g., Refs.~\onlinecite{Zap1,Glaz} for details). In particular, it allows one to make ESR measurements in a wide range of frequencies extending far beyond the bandwidth of optical detectors (up to THz) with no special microwave equipment, to identify statistics of the spin carriers, to perform three-dimensional tomographic measurements with a fairly high spatial resolution, to penetrate inside inhomogeneously broadened bands. 
A highly important advancement of the SNS was related to an advent of the Fourier-transform-based technique of spin noise (SN) detection, which has dramatically improved the sensitivity of the method~\cite{Roemer}, making it possible, for instance, to measure spin noise of a single spin \cite{singlehole} or to detect nuclear spin fluctuations~\cite{oest:nsn}.
Thanks to all these opportunities and in combination with almost nonperturbative character of the measuring procedure, the SNS has made it possible to perform many fascinating experiments, see Ref.~\onlinecite{oest:rev} for review, and thus has turned, nowadays, into a highly useful and, in many respects, unique method of research in the field of magnetic resonance and spin dynamics in semiconductors.

Application of the SNS to semiconductors and, in particular, to semiconductor nanostructures such as quantum wells and quantum dots requires an increase of the polarimetric sensitivity of the method. It can be achieved either by direct increase of the probe beam power~\cite{Zap82,Zap2} or by placing the sample inside a Fabry-Perot cavity~\cite{AKav}. In these studies, it became clear that, in practice, SNS cannot remain perfectly nonperturbative. This fact, although hinders nonperturbative measurements, may be useful for studying dynamics of non-equilibrium spin systems and thus provide the basis of the nonlinear SNS. Examples of the nonlinear SNS have already been given by the experimental studies~\cite{Roemer,Zap2,Crooker1,Huang,Polt1} in which the effects of probe beam intensity on the detected spin noise spectra were noticed. Especially spectacular example of the nonlinear SNS has been reported in the recent work~\cite{Ryzhov15}, where the elliptically polarized probe beam in the region of nominal transparency of the sample ($n$-doped GaAs) has made it possible 
to optically orient the host lattice nuclear spin system of the semiconductor in the Voigt geometry. This is possible if additional fields, besides external one, affect either electron or nuclear spin systems~\cite{Fleisher}.

It is known that the electron spin noise (SN) spectra at arbitrary orientation of the external magnetic field generally reveal two components~\cite{Zap1,gi2012:noise}. One of them is centered at zero frequency and reflects fluctuations of the longitudinal (with respect to the applied magnetic field) magnetization, while the other, at Larmor frequency, results from fluctuations of the transverse magnetization. Relative contributions of these two components are controlled by mutual orientation of the light beam and magnetic field and therefore can be used to monitor direction of the effective  magnetic field acting upon the spin system under particular  
experimental conditions.  

In this paper, we use the spin noise spectroscopy to study mechanisms of nonlinear interaction of elliptically polarized probe light of sub-bandgap energy with electron spin system of the $n$-GaAs crystal. The main attention is devoted to an analysis of properties of the effective field, termed hereafter as \emph{optical} magnetic field, created by circularly polarized component of the probe beam and revealed in a highly spectacular form in the spin noise spectrum of the sample. 
We discuss possible origin of the optical field and demonstrate that it
most likely results from the probe-helicity-dependent optical Stark effect relevant  at high power densities of the electromagnetic field inside the microcavity even for the medium transparent at the appropriate wavelength. A theory of such an effect for bulk microcavities is developed.

\section{Sample and experimental setup}\label{ec:setup}

In our studies we used sample T695, which, like the one used in Ref.~\onlinecite{Ryzhov15}, was a 3$\lambda$/2-layer of $n$-type GaAs embedded into a graded high-Q (quality factor $Q\sim 10^4$) microcavity.  Bragg mirrors of the sample were made of GaAs/AlAs, rather than AlGaAs/GaAs~\cite{Ryzhov15}, pairs of layers. This allowed us to obtain approximately the same $Q$-value of the cavity as in Ref.~\onlinecite{Ryzhov15} with smaller number of layers: 25 and 17 pairs for the first and second mirrors (starting from the side of substrate). The room-temperature doping $\sim 4\times10^{16}$ cm$^{-3}$ is slightly above the point of  insulator-to-metal transition of the GaAs and provides the longest spin relaxation times~\cite{Kikkawa,Dzhioev,Romer}. The detected spin noise is provided by electrons with energies in the vicinity of the Fermi level since (i) the magnitude of the SN grows linearly with temperature (the behavior typical for free electrons in GaAs \cite{Crooker1}) and (ii) the precession peak in the SN spectrum corresponds to the electron Larmor frequency in GaAs with the $g$-factor $|g|\approx 0.44$ (see below for details).

Magnetic resonance in the Kerr-rotation noise spectrum was detected in the conventional Voigt configuration (see, e.g., Refs.~\onlinecite{Polt1,Polt2}). Magnetic fields  up to 0.7~T and low temperatures down to 3~K were created by the Montana Cryostation system with magneto-optic module. The output emission of a continuous wave Ti:Sapphire laser ``T\&D-Scan'' tuned to the cavity photon mode (usually close to $\lambda = 833$ nm) was used as a probe. The beam was focused to a spot of about $ 20$~$\mu$m in diameter on the sample surface. Polarization noise of the beam, reflected from the sample in nearly autocollimation geometry, was detected by a balanced photoreceiver with the bandwidth $\sim $ 220 MHz and processed with a broadband FFT spectrum analyzer. As a result, the measured signal $\mathcal S(\omega)$ was proportional to the frequency spectrum of Kerr rotation angle fluctuation $(\delta \vartheta_K^2)_\omega$. Since $\delta \vartheta_K$ and electron spin fluctuation $\delta S_z$ are directly proportional, we obtain
	\begin{equation}
		\label{SNS}
		\mathcal S(\omega) \propto (\delta S_z^2)_\omega.
	\end{equation}
	Hereafter, we use the coordinate frame with the $z$- and $x$-axes being, respectively, the light propagation direction and the axis of the transverse magnetic field.

As compared with standard SNS configuration, the experimental set-up was modified to simultaneously use the probe beam as the circularly polarized pump. A quarter-wave plate introduced into the linearly polarized beam for this purpose allowed us to control  the light ellipticity. Besides, in some cases, we applied to the sample an additional magnetic field of several tens of mT directed along the light beam propagation axis to increase the pumping efficiency. For this purpose, we usually employed a permanent magnet.

\section{Phenomenology of the nonlinear response}\label{sec:exper}

We emphasize that the wavelength of the probe $\lambda\gtrsim 833$~nm beam corresponds to the region of nominal transparency of the sample, $\lambda> \lambda_g$, where $\lambda_g \approx 820$~nm  is the wavelength corresponding to the fundamental absorption edge of GaAs. Thus, the nonlinear effects in the SNS of the microcavity sample observed at relatively low intensities of the probe beam arise due to high power density of the light field inside the high-Q microcavity. 

 \begin{figure}
\includegraphics[width=0.9\columnwidth,clip]{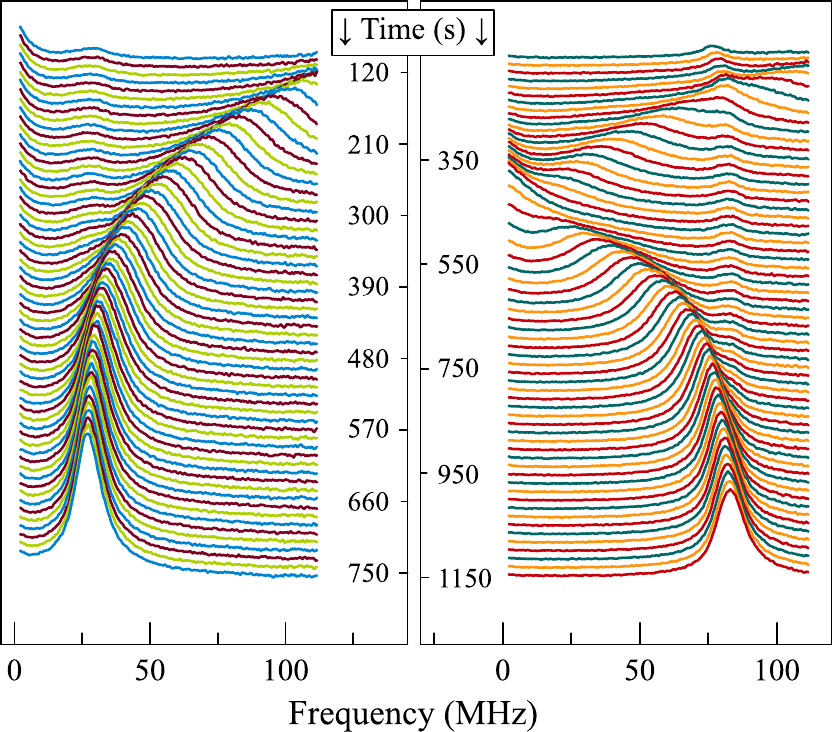}
\caption{Time evolution of the SN spectrum after pumping the sample, for several minutes, by circularly polarized probe beam in a longitudinal magnetic field. The two panels correspond to different signs of circular polarization for the same sign of the applied longitudinal field. Plots are shifted in vertical direction for visual convenience. Accumulation time for each curve is $\sim$ 1 s. }
\label{fig:1}
\end{figure}

There are two types of the light-induced nonlinear effects in this system: those with temporaly retarded response related to dynamic nuclear spin polarization and their spin dynamics uncovered in Ref.~\onlinecite{Ryzhov15} and those appearing instantaneously within accessible time resolution, see below, and related, in particular, to optical magnetic field. 

\subsection {Retarded response}

Effects of the first type, as has been reported in \cite{Ryzhov15}, are revealed in the most pronounced way after keeping the sample in the longitudinal magnetic field in the presence of circularly polarized probe beam. Under these conditions, nuclear system of the sample in the microcavity is efficiently oriented optically, and dynamics of the relaxing nuclear system evidenced via Overhauser field-induced time-dependent shift of the SN peak can be directly observed in temporal dynamics of the SN spectrum in the Voigt geometry. Figure \ref{fig:1} shows example of such dynamics obtained in this way with our sample. The slow (retarded) response of the SN spectrum is observed as a drift of the precession peak in the field of relaxing nuclear spin system with characteristic times $\sim$ 150 \ldots 200 s.  Presented results well correlate with those of Ref. \cite{Ryzhov15}, obtained on another sample, differing from them by inessential for our purposes numerical values of the SN power and width of the electron spin resonance peak. The right and left panels of Fig.~\ref{fig:1} correspond to the signs of circular polarization and longitudinal magnetic field being the same or opposite. The faint peaks in the spectrum, which are practically time-independent are most probably related to a small unintentional doping of the first few layers in the Bragg mirrors. The nuclear spins in the barriers are weakly polarized and their spin dynamics is expected to be faster, hence the spin precession of these electrons is almost unaffected by the Overhauser field. These effects are beyond the scope of the present paper.

\begin{figure}[t]
	\includegraphics[width=0.9\columnwidth,clip]{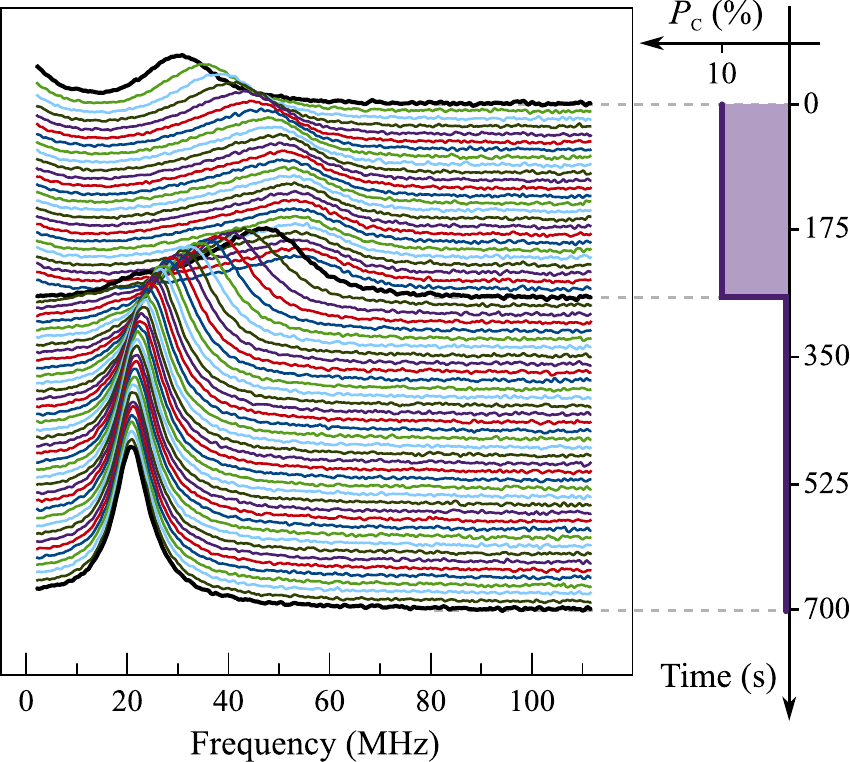}
	\caption{Response of the SN spectrum to switching the probe beam ellipticity on and off. Right panel shows the time dependence of the circular polarization degree of the probe beam.}
	\label{fig:2}
\end{figure}

The effect of the retarded optical response, associated with nuclear spin dynamics, can be also observed in the Voigt geometry with no longitudinal magnetic field just by switching probe beam polarization to elliptical and back to linear. In this case, see Fig.~\ref{fig:2}, we can observe dynamics of both nuclear spin polarization and nuclear spin relaxation.

\begin{figure}[t]
	\includegraphics[width=0.9\columnwidth,clip]{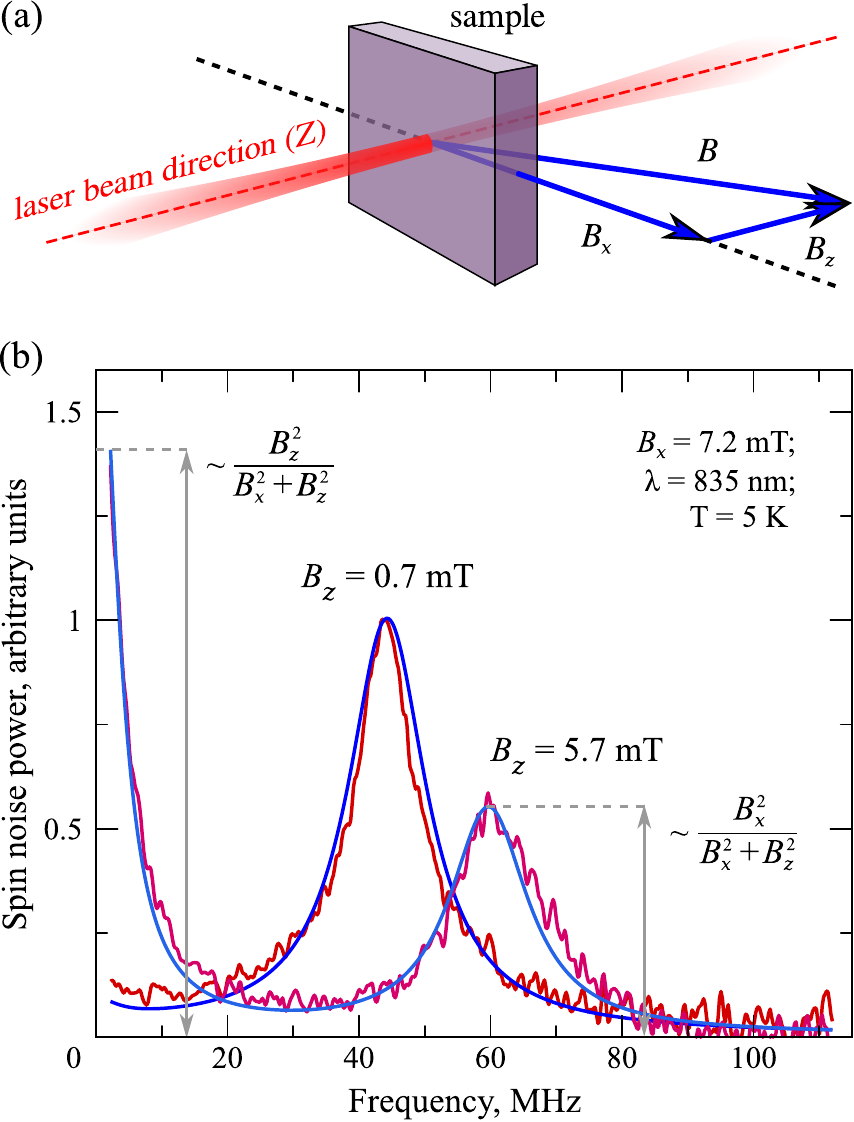}
	\caption{{(a)} Geometry of the experiment and vector diagram illustrating magnetic fields detected by the fluctuating SN system. $B_x$ is the external {\it transverse} magnetic field and $B_z$ is the {\it longitudinal} magnetic field, which may be created either by external permanent magnet or by circularly polarized probe beam.
	{(b) SN spectra} obtained at fixed transverse magnetic field $B_x$ for the longitudinal field $B_z$ being on and off. Blue lines 
	show the fit of the experimental data, see the text for details (0.7 mT, indicated at the figure, is the residual longitudinal magnetic field derived from the fitting).   	} 
	\label{fig:3}
\end{figure}

\subsection{Instantaneous response}

In addition to the retarded response, already addressed both theoretically~\cite{smirnov15} and experimentally~\cite{Ryzhov15} and lying beyond the scope of this paper, the time-resolved SN spectrum contains a stepwise jump, occuring simultaneously with switching ellipticity of the probe beam ($P_{c}$) on and off. This instantaneous response is revealed as appearance or vanishing of the peak at $\omega=0$ with simultaneous shift of the precession peak to higher or lower frequencies, see Fig.~\ref{fig:2}. Exactly the same effect can be achieved in experiments with additional magnetic field if $B_z$ is abruptly switched on or off. Therefore, switching of the ellipticity $P_{c}$ is equivalent  to an instantaneous change of the longitudinal magnetic field acting upon the electron spins. Experimentally, the retarded and instantaneous effects can be easily separated by making rapid measurements with accumulation times much shorter that that of nuclear spin relaxation. 

Time resolution of our setup ($\sim 10$~$\mu$s) did not allow us to address the timescale of instantaneous response and, thus, to find out whether this effect is formed on the timescale relevant for electron spin dynamics ($\sim$~ns) or on subpicosecond scale relevant for optical transitions. Hence, to prove that the instant modifications of the SN spectrum result from the effective optical field acting on electron spin we rely on magnetometric capacity of the SNS. It is based on the fact that the magnetic field components aligned across ($B_x$) and along ($B_z$) the light beam propagation contribute essentially different to the SN spectrum. 

\subsection{The spin noise-based magnetometry and ``optical field''}

We recall that the SN spectrum in Eq.~\eqref{SNS} in the presence of a magnetic field, containing, along with transverse component $B_x$, the longitudinal one $B_z$ takes the form~\cite{gi2012:noise}:
	\begin{multline}
		\label{eSNS}
		(\delta S_z^2)_\omega = \frac{\pi n_e}{4}\left[\Delta_{T_2}(\omega - \Omega) + \Delta_{T_2}(\omega + \Omega) \right]\cos^2{\varphi}\\
		+ \frac{\pi n_e}{2}\Delta_{T_1}(\omega)\sin^2{\varphi}.
	\end{multline}
	Here, $n_e$ is the density of fluctuating spins, for free electrons $n_e = \int \mathcal D(\varepsilon)f(\varepsilon)[1-f(\varepsilon)] d\varepsilon$, with $\mathcal D(\varepsilon)$ being the density of states including contributions from two spin  branches and $f(\varepsilon)$ being the distribution function~\cite{Crooker1,ultra:fermi,giri}, $\varphi$ is the angle between the magnetic field and $x$-axis, $\tan{\varphi} = B_z/B_x$, the electron spin precession frequency is $\Omega = g\mu_B B/\hbar$, $g$ is the electron $g$-factor, $\mu_B$ is the Bohr magneton, $B=\sqrt{B_x^2+B_z^2}$, 
	\[
	\Delta_T(x) = \frac{1}{\pi} \frac{T}{1+x^2T^2},
	\]
$T_1$ is the longitudinal electron-spin relaxation time, and $T_2$ is the transverse electron spin relaxation time. In Eq.~\eqref{eSNS}, the equilibrium magnetization of electrons in the field $\bm B$, as well as optically induced spin polarization, are disregared.

Formula~\eqref{eSNS} describes a known result that, in the general case, the SN spectrum consists of two components: The one centered at magnetic resonance frequency $\Omega$, often referred to as {\it magnetic} peak and the one centered at zero frequency and called {\it nonmagnetic}. Spectral position of the magnetic component $\Omega$ is governed by the absolute value of the total magnetic field $\bm B$, Fig.~\ref{fig:3}, while the relative magnitude of the nonmagnetic component contains information about orientation of $\bm B$ and about its longitudinal component $B_z$.

This type of behavior is illustrated by the two experimental SN spectra in Fig.~\ref{fig:3}(b) obtained, respectively, in pure Voigt geometry and in the presence of an additional static magnetic field $\bm {B}_z$ created with a permanent magnet.  One can see both the appearance of the nonmagnetic (zero-frequency) component in the presence of $B_z$ and a shift of the magnetic peak to higher frequencies. By fitting experimental data with Eq.~\eqref{eSNS}, one obtains corresponding values of the external magnetic field thus confirming reliability of the method. Extracted spin relaxation times are $T_1 \approx 44$~ns and $T_2\approx 22$~ns, in reasonable agreement with literature data for similar doping level of the bulk GaAs~\cite{Dzhioev}.

\begin{figure}[tbp]
	\includegraphics[width=0.9\columnwidth,clip]{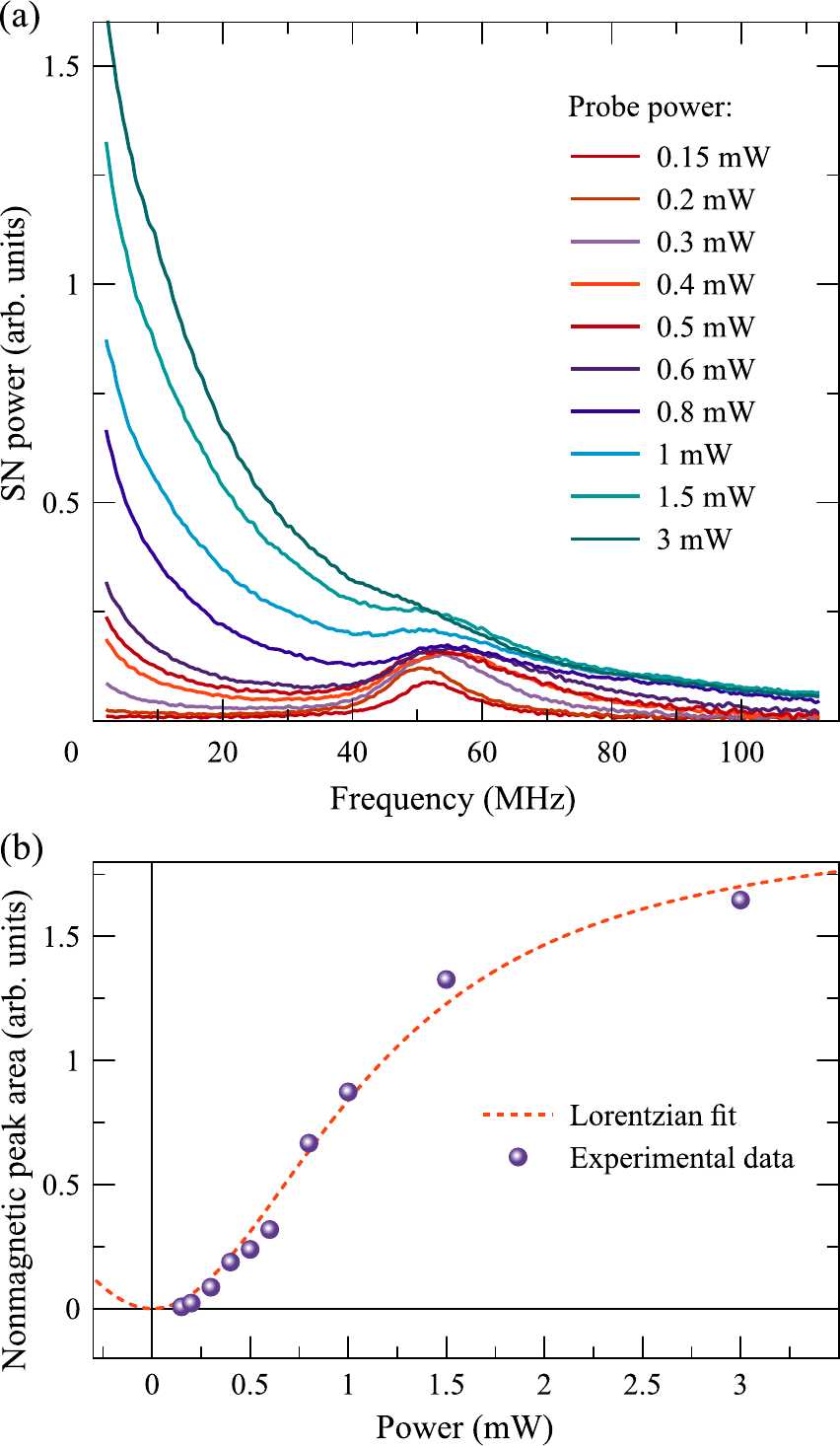}
	\caption{(a) SN spectra recorded in the presence of the elliptically polarized probe beam ($P_c\approx 20$~\%) of different intensity and (b) dependence of nonmagnetic component seen in panel (a) on the light beam power, Dashed line is the fit after Eq.~\eqref{A0A}.}
	\label{fig:4}
\end{figure}

\begin{figure}
	\includegraphics[width=0.9\columnwidth,clip]{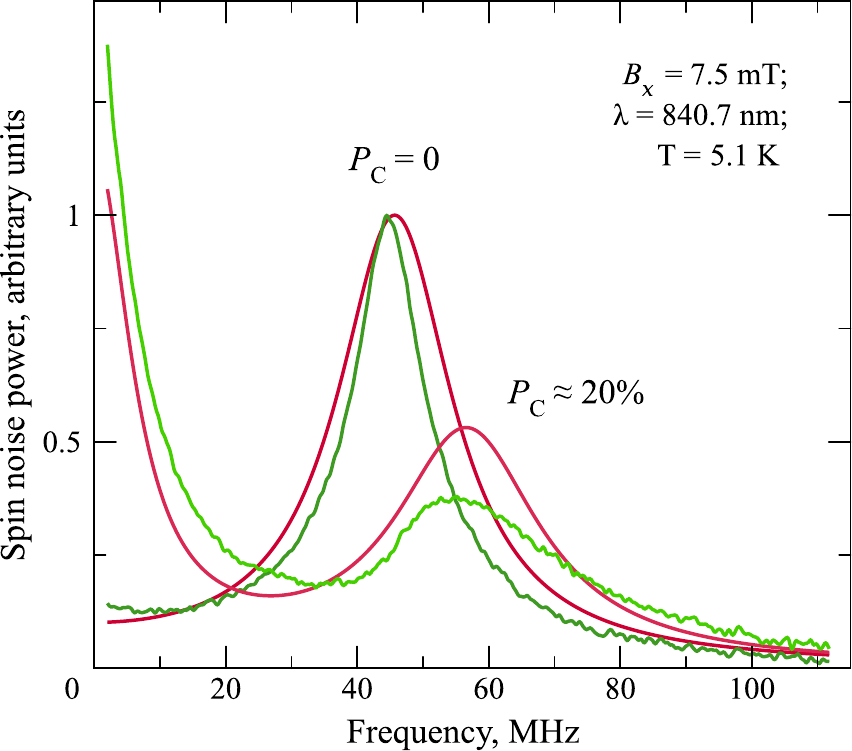}
	\caption{Comparison of experimental shape of the SN spectrum detected in the linearly and elliptically polarized light (green curves) with the results of fitting (red curves).}
	\label{fig:5}
\end{figure}

These magnetometric abilities of the SNS are revealed in a much more spectacular and nontrivial way in our present experiments with elliptically polarized probe.  Using the SN-based magnetometry, we observed an effective magnetic field (\emph{optical  field}) produced by circularly polarized probe beam in the $n$-GaAs microcavity.

Experimental results presented in Fig.~\ref{fig:4} illustrate what happens when we use an elliptically polarized probe beam instead of the beam polarized linearly. Such a configuration allows one to measure the SN spectrum in the light field with a specified helicity and, thus, to use the light beam both as a linearly polarized probe and as a circularly polarized pump.  
These measurements were performed fast enough, so that slow processes of nuclear spin orientation and relaxation could not contribute to results of the measurements.  
One can see  that, with increasing intensity of the elliptically polarized probe beam, there arises the zero-frequency component of the SN spectrum. It serves as a direct evidence of  presence of longitudinal component  of the effective magnetic field acting on electron spins. At the same time, the magnetic component shifts to higher frequencies. Again, the experimental data can be fitted by Eq.~\eqref{eSNS} with reasonable accuracy. Insignifical discrepacy between the experimental and calculated data may result from non-uniformity of the light spot. Example of such a comparison is shown in Fig.~\ref{fig:5}. Using Eq. \eqref{eSNS}, we can roughly estimate this field as $B^{opt}_z\sim 5$ mT for the probe beam power $\sim 0.5$ mW and circular polarization degree $P_{c} \approx 20\%$.


Phenomenological features of the instantaneous response demonstrate fairly convincingly that it is related to light-induced modification of the magnetic field acting upon the fluctuating electron spins, rather than to light-induced changes of the electron-spin properties.  The properties and nature of this {\it optical} field $\bm B^{opt}$ are discussed in the next section.

\section{Nature of the optical field}\label{sec:nature}

The optical field revealed in our SNS experiments shows certain features that look nontrivial and deserve special attention. In this section, we discuss these specific features and propose a theoretical model for the optical field.

\subsection{Discussion}\label{sec:exp_facts}

First of all, the optical field is mainly directed along the light propagation ($z$) axis, see Sec.~\ref{sec:exper}, and changes its sign with reversal of the probe beam helicity. The latter was checked by compensating the optical field by a longitudinal field of a permanent magnet.

Second characteristic feature of the optical field is that there is no noticeable time delay between the  stepwise jump in polarization of the probe and  corresponding  change of the SN signal. As has been shown in our additional measurements, the response of the SN  signal to polarization of the probe, within our instrumental precision ($\sim 10 \mu$s), was instantaneous. In any case, the effect of nuclear spin system in formation of optical field $\bm B^{opt}$ can be safely disregarded.

Thus, there are basically two options for the origin of the optical field. It could be either caused by (i) some electron spin system or by (ii) electromagnetic radiation itself. Let us examine these options in more detail.

We note that although the interaction between electrons is spin dependent, the electrons probed by the SNS cannot be responsible for the optical field. This follows from the Larmor theorem which states that the spin resonance frequency is not renormalized by the electron-electron interactions~\cite{Larmor}. Deviations from the Larmor theorem can be observable only in low-dimensional semiconductor systems with spin-anisotropic interactions and, as a rule, are minor~\cite{Larmor:v}.  Hence, there could only be a contribution from some other electron ensemble, i.e., from electrons localized at donor pairs~\cite{giri1} or from photogenerated holes.

\begin{figure}
	\includegraphics[width=0.9\columnwidth,clip]{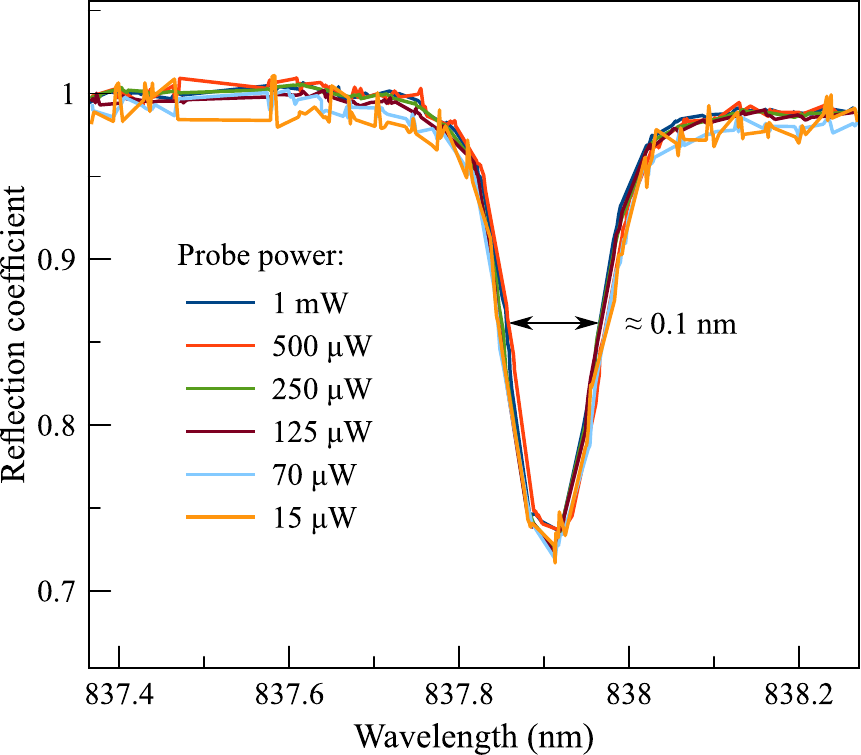}
	\caption{Experimental plots illustrating independence of the cavity resonance with and, accordingly, the $Q$-factor on the probe beam power.}
	\label{fig:6}
\end{figure}

However, the optical magnetic field is produced, in our experiments, by the light with the photon energies below the GaAs bandgap, which corresponds to the region of nominal transparency of the system. We note also that the microcavity $Q$-factor does not show any noticeable changes with the light power,~Fig.~\ref{fig:6}.
Although some absorption of light should take place to enable optical orientation of electron spins and dynamic nuclear polarization~\cite{Ryzhov15,smirnov15,abragam,opt_or_dp}, two additional spin systems  mentioned above (localized electrons and photoholes) have very short spin lifetimes, typically $\lesssim 10$~ps~\cite{kkavokin}, as compared with $\gtrsim 10$~ns spin lifetimes for free electrons in our sample, and, thus, spin polarization of these spin systems should be negligible. Even if there is some electronic spin sub-system, in our sample, with sufficiently long spin lifetime ($\tau_s \gtrsim 10$~ns) to provide sizeable polarization, the external transverse field $B_x$ should suppress its spin $z$-component due to Hanle effect and, at the same time, give rise to the other transverse component $B_y$.



The experiment has shown, however, that this is not the case: optical field $B_z$ obtained from the fitting of data is independent on $B_x$. Figure~\ref{fig:4}(b) presents experimental dependence of the nonmagnetic peak area on intensity of the {\it elliptically} polarized probe beam ($P_c \approx 20\%$) at a fixed value of the external transverse magnetic field $B_x $.  This dependence completely coincides with our theoretical predictions for the optical field, see below, and, thus, makes it possible to quantitatively evaluate, with sufficiently high accuracy, magnitude of this field.  

Indeed, it follows from Eq.~\eqref{eSNS} that the ratio of the area of nonmagnetic component $A_0$ of the SN spectrum to the total area $A$ is simply given by~\cite{note:A0}
\begin{equation}
\label{A0A}
\frac{A_0}{A} = \sin^2{\varphi} = \frac{B_z^2}{B_x^2+B_z^2}.
\end{equation} 
Taking into account that the optical field is parallel to $z$ axis, and that its magnitude is proportional to both incident light helicity, $P_c$, and intensity, $I$
\begin{equation}
\label{Bopt:phen}
B_z = B^{opt} = P_c \varkappa_c I=P_c \mathcal K_c  W ,
\end{equation} 
where $W = \mathcal A I$ is the total light power in the illuminated spot, with $\mathcal A$ being the area  the spot, $\varkappa_c$ and $\mathcal K_c= \varkappa_c/\mathcal A$ are constants.
Making use of the fact that $B_x$ is just the external field and fitting the data in Fig.~\ref{fig:4}(b) by Eq.~\eqref{A0A} we obtain  $P_c \mathcal K_c^{\rm exper}\approx ( 8.6 \pm 1.5)$~mT/mW~\cite{sign1} or 
\begin{equation}
\label{Kc:exper}
\mathcal K_c^{\rm exper} \approx( 43 \pm 7)~\frac{\mbox{mT}}{\mbox{mW}}.
\end{equation}
Thus, we can make an important conclusion, that neither the magnitude, nor the direction of the optical field are affected by the transverse field $B_x$ (at least for $B_x<20$~mT).

\subsection{Model for the optical field formation}\label{sec:theor_model}
	
Irradiation of the quantum systems in the transparency region results in the renormalization of the electron energy spectrum~\cite{floq0}. This class of effects is known as \emph{ac} or \emph{dynamic} Stark effects, see Ref.~\onlinecite{suss} for brief review. Particularly, circularly polarized radiation results in the effective Zeeman splitting of electron spin states, sometimes referred to as \emph{ac} Zeeman effect~\cite{china}. The effect of circularly polarized light can be thus considered as the generation of the effective \emph{optical} magnetic field, just like the propagation of circularly polarized light in transparent media results in its magnetization, termed as inverse Faraday effect~\cite{pitaevskii,PhysRev.143.574}. By symmetry reasons in cubic crystals like GaAs ($T_d$ point symmetry group) this effective magnetic field can be presented in the linear in the light intensity
\begin{equation}
 I = \frac{c}{2\pi n_b} |E|^2
\end{equation} 
regime as
\begin{equation}
\label{Bopt:sym}
\bm B^{opt} = \bm n \ P_c  \varkappa_0 I.
\end{equation}
Here $n_b$ is the background refractive index and $E$ is the amplitude of the field, $\bm n$ is the unit vector in the direction of light propagation and $\varkappa_0$ is a coefficient, which, in general depends on the light intensity, 
Symmetry properties of the optical field $\bm B^{opt}$ introduced in this way directly correspond to the experimental observations reported above.

To evaluate the optical field we use the perturbation theory and present an effective Hamiltonian of the electron in crystal under illumination by light with the frequency $\omega$ in the second order in the light field amplitude $\bm E_0$ as~\cite{PhysRev.143.574,birpikus}
\begin{equation}
\label{Heff}
\delta \mathcal H_{n'n} = -\sum_{\alpha\beta} \chi_{n'n}^{\alpha\beta} E_{0\alpha}^*E_{0\beta},
\end{equation}
where $\alpha, \beta$ are the Cartesian components,
\begin{multline}
\label{chi}
\chi_{n'n}^{\alpha\beta} = -\frac{e^2}{m_0\omega^2}  \\
\times \left[\delta_{nn'}\delta_{\alpha\beta} + \frac{1}{m_0\hbar} \sum_{m\ne n,n'} \left(\frac{p^\beta_{n'm}p^\alpha_{mn}}{\omega_{mn}-\omega} +\frac{p^\alpha_{n'm}p^\beta_{mn}}{\omega_{mn}+\omega} \right) \right],
\end{multline}
$n, n',m,\ldots$ enumerate the states (including band index, wavevector and spin), $\omega_{mn} = (E_m - E_n)/\hbar$ are the transition frequencies, $e$ is the electron charge, $m_0$ is the free electron mass, and $p_{mn}^\alpha$ are corresponding components of electron momentum operator taken between the states $m$ and $n$.
For the conduction band states with the wavevector $\bm k$ the effective spin Hamiltonian ($s,s'=\pm 1/2$) can be written as
\begin{multline}
\label{Heff:s}
\delta \mathcal H_{s's} = -\chi_{\bm k,s';\bm k,s}^{\alpha\beta} E_{0,\alpha}^*E_{0\beta} \\ = -\chi_{\bm k,s';\bm k,s}^{(s),\alpha\beta}\frac{E_\alpha^*E_\beta+E_\alpha E_\beta^*}{2}-\chi_{\bm k,s';\bm k,s}^{(a),\alpha\beta}\frac{E_\alpha^*E_\beta-E_\alpha E_\beta^*}{2} ,
\end{multline}
where to shorten the notations we have omitted the summation over repeated subscripts and introduced symmetric and antisymmetric with respect to the permutations of $\alpha$ and $\beta$ components of the tensor $\chi$:
\[
\chi_{\bm k,s';\bm k,s}^{(s),\alpha\beta}=\frac{\chi_{\bm k,s';\bm k,s}^{\alpha\beta}+\chi_{\bm k,s';\bm k,s}^{\beta\alpha}}{2},
\]
\[
\chi_{\bm k,s';\bm k,s}^{(a),\alpha\beta}=\frac{\chi_{\bm k,s';\bm k,s}^{\alpha\beta}-\chi_{\bm k,s';\bm k,s}^{\beta\alpha}}{2}.
\]
Last term in Eq.~\eqref{Heff:s} with $\chi_{\bm k,s';\bm k,s}^{(a),\alpha\beta}$ is sensitive to the helicity of light and in what follows we focus on this term only.

For the light propagating along $z$ axis the effective Hamiltonian, Eq.~\eqref{Heff:s}, can be recast in form of Zeeman splitting
\begin{equation}
\label{Heff:circ}
\delta \hat{\mathcal H} = \frac{1}{2}g \mu_B B_{z}^{opt} \hat{\sigma}_z,
\end{equation}
where $\hat\sigma_z$ is the $z$ Pauli matrix, with the optical field in the form of Eq.~\eqref{Bopt:sym} with 
\begin{equation}
\label{B}
\varkappa_0 = -\frac{{4}\pi \hbar^2 e^2 p_{\rm cv}^2{n_b}}{3c g \mu_B m_0^2 E_g^2\delta}.
\end{equation}
Here we used resonant approximation and, correspondingly, omitted in Eq.~\eqref{chi}, the first and last terms,
$n_b$ is the background refractive index, $p_{\rm cv}$ is the interband momentum matrix element, and the detuning $\delta = E_g - \hbar\omega$.
Taking $E_g=1.5$~eV, $n_b=3.66$,  $p_{\rm cv}=1.5\times 10^{-19}$~gm$\cdot$cm/s~\cite{vurgaftman02}, $g=-0.44$ one has
\[
\varkappa_0 \approx {\frac{8.5\times 10^{-5}}{(\delta/\mbox{meV})}} \frac{\mbox{mT}}{\mbox{mW/cm}^2}.
\]

\begin{figure}
 \includegraphics[width=\linewidth]{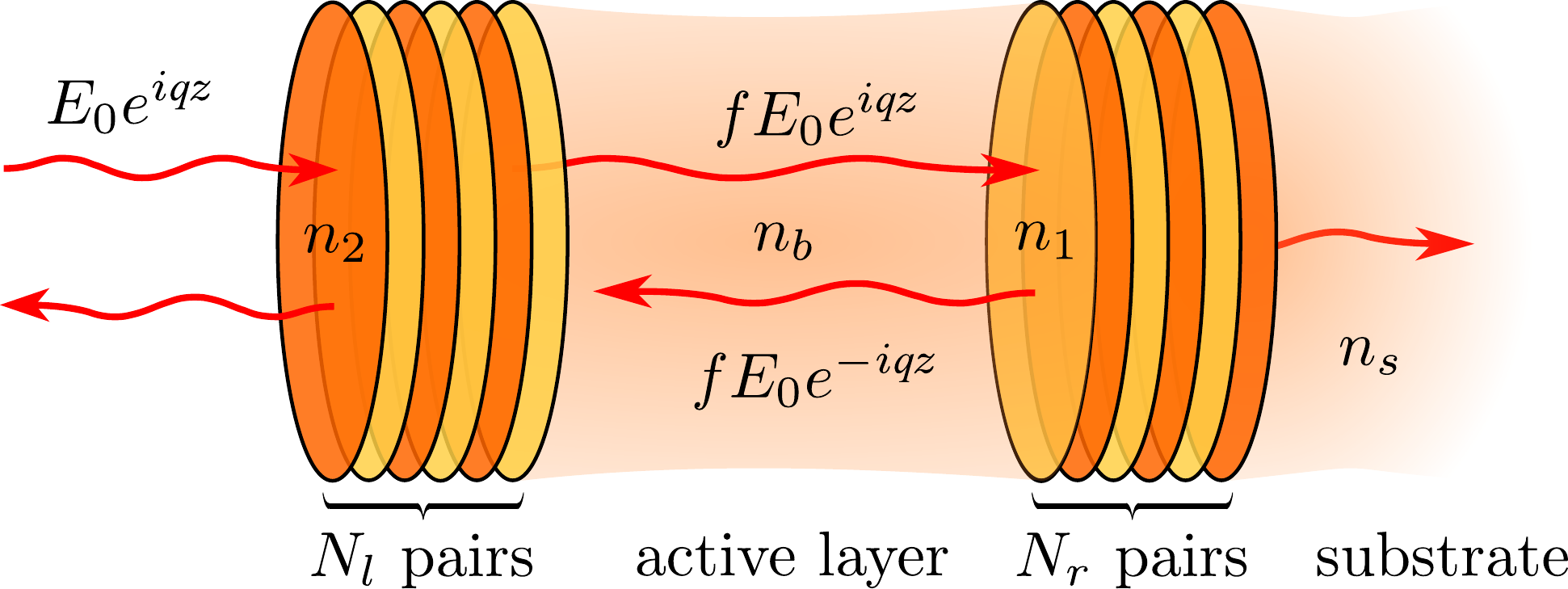}
 \caption{The sketch of the cavity and the notations. Please replace $k\to q$, since $\bm k$ stands for electron wavevector and $\bm q$ is the radiation wavevector.}
 \label{fig:7}
\end{figure}

In order to compare this result with experiment one has to take into account the enhancement of light intensity inside the cavity. To that end we assume that, like in experiment, the incident radiation frequency is in resonance with the cavity mode and represent the field inside the cavity in the form of two waves propagating to the left and to the right, Fig.~\ref{fig:7}:
\begin{equation}
\label{field}
\bm E = {f}
\bm E_0\left( e^{\mathrm i q z} + e^{-\mathrm i q z}\right), 
\end{equation}
where $\bm E_0$ is the amplitude of the field outside the cavity, $\bm q$ is the light wavevector inside the active GaAs layer, the light is assumed to be incident on the structure from the left and $z=0$ refers to the left border of the active media ($n$-GaAs layer). The enhancement factor
\begin{equation}
\label{enh}
f = \sqrt{\frac{1-R_l}{n_b}}\left(1-\frac{R_l+R_r}{2}\right)^{-1}
\end{equation}
where
$R_{l,r}$ ($1-R_{l,r} \ll 1$) are the intensity-related reflection coefficients for the light incident from the active media on the left and right mirrors, respectively.
It follows from Ref.~\onlinecite{ivchenko05a} that
\begin{equation}
\label{T}
R_l=1-\frac{4}{n_b}\left(\frac{n_1}{n_2}\right)^{2N_l}, \quad  R_r=1-4\frac{n_s}{n_b}\left(\frac{n_1}{n_2}\right)^{2N_r},
\end{equation}
where $n_s$ refraction index of the substrate, $n_1$ and $n_2$ are the refractive indices of the layers in Bragg mirrors ($n_1$ corresponds is the layers adjacent to the cavity), $N_l$ and $N_r$ are the number of layer pairs for the left and right mirrors, respectively, see Fig.~\ref{fig:7} for details.
We present also the quality factor of the $3\lambda/2$-cavity $Q$ defined as $Q=\bar\omega/(2\bar\gamma)$, where $\bar\omega$ and $\bar\gamma$ are the real and imaginary parts of the cavity resonance frequency:
\begin{equation}
\label{Q}
 Q={\frac{\pi}{n_b}\left(\frac{n_1n_2}{n_2-n_1}+3n_b\right)\left(1-\frac{R_l+R_r}{2}\right)^{-1}}.
\end{equation}

Since the number of Bragg mirrors from the side of the substrate is considerably larger than from the side of the surface, $N_r>N_l$, Eqs.~\eqref{T} yield $1-R_r\ll1-R_l$. So for the estimations one can simply put $R_r=1$. In this way we obtain the relation
\begin{equation}
 f^2=\frac{2Q}{\pi}\left(\frac{n_1n_2}{n_2-n_1}+3n_b\right)^{-1}.
 \label{f2Q}
\end{equation} 
Due to the roughnesses of the interfaces the $Q$-factor of the microcavity may be smaller than the value obtained from Eq.~\eqref{Q}. Thus it is preferable to estimate the factor $f$ from Eq.~\eqref{f2Q} using the experimental value of $Q$, than directly using Eq.~\eqref{enh}.

For the parameters of the structure under study: $n_1=2.98$, $n_2=n_s=3.66$, $N_l=17$, $N_r=25$, $Q=8700$ the Eq.~\eqref{f2Q} gives $f^2=205$. Note that substantial difference between $Q$ and $f^2$ results from the effective field penetration into Bragg mirrors. Hence, $\mathcal K_c$ in Eq.~\eqref{Bopt:phen} reads
\begin{equation}
\label{Kc:theor}
\mathcal K_c^{\rm theor}  = \frac{2f^2\varkappa_0}{n_b\mathcal A} \approx 55~\frac{\mbox{mT}}{\mbox{mW}},
\end{equation}
for the detuning $\delta=25$~meV and probe area $\mathcal A=\pi r_0^2$, $r_0 = 15$~$\mu$m. We note good agreement both in the magnitude and in the sign between this simple estimation and the experimental value, Eq.~\eqref{Kc:exper}. 


The formation mechanism of the optical field discussed here is related to the renormalization of the electron energy spectrum of bulk GaAs in the field of elliptically polarized light wave and the enhancement of the electromagnetic field intensity in microcavities. Besides, semiconductor microcavities demonstrate a variety of specific polarization-dependent linear and nonlinear effects~\cite{oshe,shelykh:rev,angular,shg}, including polarization conversion, self-induced Larmor precession, second harmonic generation, spin to angular momentum conversion, etc. Each of these effects may be of particular importance for the optical field induction in the strong coupling regime. The finding of a microscopic mechanism responsible for the optical field formation and nuclear spin pumping in microcavities constitute a fascinating new research problem that will be in scope of our further studies.

\section{Conclusion}

In this paper we used spin noise based magnetometry to reveal the ``optical'' magnetic field, an effective magnetic field acting on electron spins caused by the circularly or elliptically polarized light. This field arises when the light propagates through the sample in the region of transparency, scales linearly with radiation intensity and reverses its direction when the light helicity is reversed. The experiments were carried out on the bulk $n$-GaAs layer embedded into a high-$Q$ microcavity, which strongly enhances the intensity of incident radiation and makes optical field pronounced. The magnetometric abilities of the spin noise spectroscopy technique allowed us to characterize this optical field quantitatively. Calculations of the field magnitude, based on the proposed model of the optical Stark effect in the field of circularly polarized light, well correlate with the experimental data.

The demonstration of the strong optical field in bulk semiconductor opens ways for coherent manipulation of the spin states like realized recently for quantum dots~\cite{control1,control2} and for realization of non-trivial spin structures similar to those discussed for two-dimensional systems~\cite{floq1,floq2,floq3}.

Besides demonstrating informative potentialities of the spin noise spectroscopy, these results, in our opinion, are important for understanding
the effect of optical nuclear orientation observed in microcavities. However, the mechanism of the efficient transfer of angular momentum from the light to nuclear spin system under highly unfavorable conditions when the pumping light beam acts upon the system in the region of its nominal transparency  remains unsolved and intriguing problem to be analyzed in more detail elsewhere. 
		

		\section*{Acknowledgements}
		
		Financial support from the Ministry of Education and Science of the Russian Federation (Contract No. 11.G34.31.0067), Saint Petersburg State University (Grant No. 11.38.213.2014), 
		Dynasty Foundation, RF President Grants SP-643.2015.5, MD-5726.2015.2, the Russian Foundation for Basic Research and the Deutsche Forschungsgemeinschaft in the frame of International Collaborative Research Center TRR 160 (project No. 15-52-12013) is acknowledged.
		AK acknowledges support from the EPSRC  Established Career Fellowship (Grant No.  RP008833).  
		The work was carried out using the equipment of SPbU Resource Center  "Nanophotonics" (photon.spbu.ru).


\begin{thebibliography}{99}
\bibitem{note:A0} We restrict integration by positive frequencies only, in which case $A_0 = \int_0^{\omega^*} (\delta S_z^2)_\omega d\omega$ with $1/T_1 \ll \omega^* \ll \Omega$.
\bibitem{sign1} The relative sign of optical field $z$-component, $B_z$, and circular polarization, $P_c$, can be established from (i) the sign of the nuclear spin temperature (determined from the temporal dependence of the electron spin precession peak position in Fig.~\ref{fig:1}, see Ref.~\cite{Ryzhov15}) and (ii) the mutual orientation of external static and optical field: $\bm B$ and $\bm B^{opt}$ are parallel in Fig.~\ref{fig:1}(a) and anti-parallel in Fig.~\ref{fig:1}(b).
\end{thebibliography}
\end{document}